\documentclass[onecolumn,preprint]{revtex4}
\usepackage{amsfonts, amssymb,amsmath} 
\usepackage{graphicx}
\usepackage{dcolumn}
\usepackage{bm}

\newcommand{\be}{\begin{equation}}
\newcommand{\ee}{\end{equation}}
\newcommand{\bea}{\begin{eqnarray}}
\newcommand{\eea}{\end{eqnarray}}

\newcommand{\ti}{\i}

\newcommand{\tro}{\"o}

\newcommand{\Sch}{Schr\tro dinger equation~}

\catcode`ð=\active \defð{\u{g}} \catcode`Ð=\active \defÐ{\u{G}}
\catcode`Ý=\active \defÝ{\. I} \catcode`ö=\active \defö{\"{o}}
\catcode`Ö=\active \defÖ{\"O} \catcode`ü=\active \defü{\"{u}}
\catcode`Ü=\active \defÜ{\"{U}} \catcode`Þ=\active \defÞ{\c{S}}
\catcode`þ=\active \defþ{\c{s}} \catcode`ý=\active \defý{{\i }}
\catcode`ç=\active \defç{\c{c}} \catcode`Ç=\active \defÇ{\c{C}}

\begin{document}


\title{A Parabolic Model for the Dimple Potentials}

\author{Melike  \c{C}\i b\i k Ayd\i n$^1$, Haydar Uncu$^1$, Co\c{s}kun Deniz$^2$}
\affiliation{$^{1}$Department of Physics, Adnan Menderes University,
Aytepe, 09100, Ayd\ti n, Turkey, \\ $^{2}$Department of Electronics
Engineering, Adnan Menderes University, Aytepe, 09100, Ayd\ti n,
Turkey}
\date{\today}
\begin{abstract}
We study truncated parabolic function and demonstrate that it is a
representation of the Dirac $\delta$ function. We also show that the
truncated parabolic function, used as a potential in the
Schr\"{o}dinger equation, has the same bound state spectrum,
tunneling and reflection amplitudes  with the Dirac $\delta$
potential as the width of the parabola approximates to zero. Dirac
$\delta$ potential is used to model dimple potentials which are
utilized to increase the phase-space density of a Bose-Einstein
condensate in a harmonic trap. We show that harmonic trap with a
$\delta$ function at the origin is a limit case of the harmonic trap
with a symmetric truncated parabolic potential around the origin.
Therefore, we propose that the truncated parabolic potential is a
better candidate for modeling the dimple potentials.
\end{abstract}
\maketitle
\section{Introduction \label{intro} }
The dimple potentials are used to increase the phase space density
of the  dilute Bose gases
\cite{Pinkse,stamper,ma,garrett,jacob,hammes}.
This increase gives rise to the the formation of a Bose-Einstein
condensate (BEC) when the initial sample is just above the critical
temperature. So, the dimple potentials provide a means of inducing
condensation without cooling \cite{garrett}. Moreover, the
improvements in the trapping techniques of cold atoms made possible
to realize Bose-Einstein condensates in low dimensions
\cite{bloch,gorlitz,ott,teo,gunter}. Thus, theoretical models of
dimple potentials in one dimension have been studied in the
literature \cite{utem,busch}. While some of these models utilize an
inverse Gaussian function others adopt Dirac $\delta$ functions to
describe dimple potentials. We will use a truncated parabolic
potential $V(x)= -U_0 (1-x^2/a^2)$ for $x\leq \vert a \vert$ to
model dimple potentials first proposed in Ref. \cite{ma} for three
dimensions. However, in this study Ma et. al. used a numeric
simulation to explain for both the loading process and the
subsequent evaporative cooling from a dimple potential.

The advantage of the parabolic model compared to Gaussian model is
that the \Sch for the truncated parabolic potential is analytically
solvable, so one obtains the thermodynamic quantities of a weakly
interacting Bose gas at least as a first approximation. Although the
\Sch for Dirac $\delta$ potentials are also analytically solvable,
one can not incorporate both the dimple depth and width together
into these models. Only the multiplication of these quantities are
used as the coupling coefficient of the Dirac $\delta$ potential
\cite{utem}. In addition to this, Dirac $\delta$ models are only
valid for tough and narrow dimples. However, experiments with
relative wide dimples are also performed \cite{garrett}. Therefore,
the parabolic model which accommodates both dimple depth and width
separately is more useful in describing dimple potentials.

As a special case of point interactions, Dirac $\delta$ potential
has been taking interest for years. Atkinson and Crater investigated
the effect of a Dirac $\delta$ to the bound states of various
potentials in one dimension \cite{attkinson}. Several authors made
further studies on this subject \cite{avakian,demiralp2,wang}. Bound
states of arbitrary but finite number of Dirac $\delta$ shells in
higher dimensions are also studied \cite{demiralp}. The resonances
of potentials with several Dirac $\delta$ are examined by Gadella
et. al. \cite{gadella,alvarez}. In physical applications, Dirac
$\delta$ potential is mainly used to describe short range effects.
For example, Erkol et. al. used $\delta$ potential for modeling
short range interactions in atomic and nuclear physics
\cite{erkol1,erkol2}. Moreover it is shown that $\delta$ potentials
can also be used to find approximate solutions for tunneling and
reflection amplitudes for different potentials \cite{sahu}.

In this study, we suggest that the truncated parabolic function
$f_{\delta}(x)=3/(4 a)\, (1-x^2/a^2) $, for $\vert x \vert \leq a$
is a representation of the Dirac $\delta$ function. Moreover, since
the \Sch of the truncated  parabolic potential  is analytically
solvable, we are able to show that the bound state spectrum,
tunneling and reflection coefficients of the truncated parabolic
potential go to those of the Dirac $\delta$ potential as the width
decreases while the depth increases. We also demonstrate that the
spectrum of the harmonic trap with a truncated parabolic potential
reduces to the spectrum of the harmonic potential with a Dirac
$\delta$ function in the same limit. So, the truncated parabolic
potential model of the dimple trap  includes all the results
obtained for the non-interacting Bose gases by the Dirac $\delta$
models for the dimple potential as a special case.

The paper is organized as follows: In section \ref{HTWPP},  we
present the solution of the Schr\"{o}dinger equation for a harmonic
trap with a truncated parabolic potential in one dimension and
obtain eigenvalue equations and eigenfunctions, for the sake of
completeness. In order to find the eigenvalues, the eigenvalue
equations are solved numerically. In section \ref{JWKB}, we apply
JWKB approximation to check the validity of numerically found
eigenvalues. In section \ref{TA}, we use the sudden perturbation
theory to find the transition amplitudes from harmonic trap to the
harmonic trap with a dimple described by a truncated parabolic
potential. In section \ref{DDPaLC}, we show that the truncated
parabolic function provides a representation of the Dirac $\delta$
function. Moreover, we also demonstrate that the spectrum, the
reflection and transmission coefficients of this potential reduces
to those of the Dirac $\delta$ potential as the width of the
truncated parabolic potential goes to zero, when the depth times
width is fixed. Finally, we conclude in Section \ref{CONC}.

\section{Harmonic Trap with a Parabolic Potential \label{HTWPP}}
In this section, we obtain the eigenvalues and the eigenfunctions of
the Hamiltonian
\be
\hat{H}= \frac{\hat{P}^2}{2m} + V(\hat{X})
\label{hamiltonian}
\ee
in one dimension where the potential function is given as
\be
V(x)= \begin{cases} \frac{1}{2} m \omega^2 x^2 & \textrm{for~} \vert
x \vert >a
\\
\frac{1}{2} m \omega^2 x^2 -U_0 \left( 1-\frac{x^2}{a^2} \right) &
\textrm{for~} \vert x \vert \leq a \; .
\end{cases}
\label{potential}
\ee
Here $U_0 >0$  and $a>0$ represent the depth and the width of the
truncated parabolic potential, respectively.

The time independent \Sch for this potential is
\be
-\frac{\hbar^2}{2m} \frac{d^2 \Psi(x)}{dx^2} + V(x) \Psi(x)= E
\Psi(x), \label{Scheq}
\ee
where $V(x)$ is given in the Eq. \eqref{potential}. For $U_0=0$, the
Hamiltonian reduces to the harmonic oscillator Hamiltonian whose
eigenvalues are $E_n=(n+1/2) \hbar \omega, \quad n=0,1,2,\ldots$ and
whose eigenfunctions are  given in terms of the Hermite polynomials
$H_n(\sqrt{m \omega/\hbar} \; x )$. However if $U_0 \neq 0 $, the
eigenvalues are no more $(n+1/2) \hbar \omega$ and the
eigenfunctions cannot be written in terms of the Hermite polynomials
$H_n(\sqrt{m \omega/\hbar} \; x ), \; n=0,1,2,\ldots $ whose degrees
are finite.

The time independent Schr\"{o}dinger equation of the Hamiltonian  in
Eq. \eqref{hamiltonian} whose potential term is given in Eq.
\eqref{potential} is different for $\vert x \vert
> a$ and for $\vert x \vert \leq a$. For $\vert x \vert > a$
the Schr\"{o}dinger equation reduces to
\be
-\frac{\hbar^2}{2m} \frac{d^2 \Psi(x)}{dx^2} + \frac{1}{2} m
\omega^2 x^2 \Psi(x)= E \Psi(x). \label{eigenvalue1}
\ee
Although the differential equation \eqref{eigenvalue1} is same as
the time independent \Sch for the harmonic oscillator potential, the
solutions for $\vert x \vert > a$ differ from harmonic oscillator
solution because of the continuity conditions on the wave functions
and their derivatives at the points $\vert x \vert =a$.

We continue by rewriting the Eq. \eqref{eigenvalue1} in terms of
dimensionless variable
\be
z=\left( \frac{m \omega}{\hbar} \right)^{1/2} x %
\label{dimenionless}
\ee
as
\be
\frac{d^2 \Psi(z)}{dz^2} +  \left( 2 \varepsilon - z^2  \right)
\Psi(z)= 0  \label{hypergeomz}
\ee
where $\varepsilon= E/(\hbar \omega)$. The general solutions of
Eq.\eqref{hypergeomz} can be written in terms of parabolic cylinder
functions \cite{lebedev}:
\bea
D_{\lambda}(z)&=& 2^{\lambda} e^{-z^2/2} \left\{ \Gamma(\frac{1}{2})
\frac{ \Phi\big( -\lambda/2,1/2;z^2
\big)}{\Gamma(\frac{1-\lambda}{2} )} +\Gamma(-\frac{1}{2}) z \frac{
\Phi\big(\, (1-\lambda)/2,3/2;z^2 \big)}{\Gamma(\frac{-\lambda}{2}
)} \right\} \nonumber
\\
D_{\lambda}(-z) &=& 2^{\lambda} e^{-z^2/2} \left\{
\Gamma(\frac{1}{2}) \frac{ \Phi\big( -\lambda/2,1/2;z^2
\big)}{\Gamma(\frac{1-\lambda}{2} )} -\Gamma(-\frac{1}{2}) z \frac{
\Phi\big(\, (1-\lambda)/2,3/2;z^2 \big)}{\Gamma(\frac{-\lambda}{2}
)} \right\}
\label{hypergeomsol}
\eea
where $\lambda=\varepsilon -1/2$, $\Gamma(x)$ is the well known
gamma function and $\Phi(\alpha,\gamma;y)$ is the confluent
hypergeometric function \cite{lebedev}. The asymptotic behavior of
$D_{\lambda}(z)$ and $D_{\lambda}(-z)$ are as follows:
\[ \lim_{z
\to \infty} D_{\lambda}(z) = 0, \; \lim_{z \to \infty}
D_{\lambda}(-z) = \infty \quad \textrm{~and~} \lim_{z \to -\infty}
D_{\lambda}(-z) = 0, \; \lim_{z \to -\infty} D_{\lambda}(z) = \infty
\, .
\]
Because $D_{\lambda}(\mp z)$  diverge at $\pm \infty$,  the wave
functions for $x<-a$ ($z < -(\frac{m \omega }{\hbar})^{1/2} a $) and
$ x>a$ ($z> (\frac{m \omega }{\hbar})^{1/2} a $) are
\bea
\psi_1 (z) &=& c_1 D_{\lambda}(-z) \label{WFML}
\\
\psi_3 (z) &= & c_4 D_{\lambda} (z) \label{WFMR}
\eea where $c_1$  and $c_4$ are constants that can be determined by
the normalization and continuity conditions.

For $ \vert x \vert \leq a$, Eq. \eqref{Scheq} takes the form
\be
-\frac{\hbar^2}{2m} \frac{d^2 \Psi(x)}{dx^2} +\left[ \frac{1}{2} m
\omega^2 x^2-U_0 \left( 1-\frac{x^2}{a^2}  \right) \right] \Psi(x)=
E \Psi(x) . \label{eigenvalue2}
\ee
Defining
\be
\omega_d= \sqrt{\omega^2+ \frac{2 U_0}{ma^2}}, \quad
\frac{E+U_0}{\hbar \omega_d}=\varepsilon_d , \quad
\lambda_d=\varepsilon_d-\frac{1}{2}
\label{severaldefs}
\ee
and changing the variable to $z_d=\sqrt{\frac{m \omega_d}{\hbar}} \;
x $, Eq. \eqref{eigenvalue2} takes a similar form with the Eq.
\eqref{hypergeomz}:

\be
\frac{d^2 \Psi(z_d)}{dz_d^2} +  \left( 2 \varepsilon_d - z_d^2
\right) \Psi(z_d)= 0. \label{hypergeomzd}
\ee
Therefore the solutions of this equation are same as Eq.
\eqref{hypergeomsol} with $z$ replaced by $z_d$ and $\lambda$
replaced by $\lambda_d$. Hence, the wave functions for $\vert x
\vert \leq a$ are
\be
\psi_2 (z) = c_2  D_{\lambda_d}(z_d)+  c_3 D_{\lambda_d} (-z_d)
\label{WFM}
\ee
where $c_2$ and $c_3$ are constants which can be found by
normalization and continuity conditions. Since the potential is an
even function of $x$, the eigenfunctions are either even or odd
(hence $c_4= \pm c_1$ in Eq. \eqref{WFMR} and $c_3= \pm c_2$ in Eq.
\eqref{WFM}):
\be
\Psi_{\lambda} (z) = \begin{cases} c_1 D_{\lambda}(-z) &
\textrm{for~} z < -(\frac{m \omega }{\hbar})^{1/2} a
\\
c_2 \left[ D_{\lambda_d}(z_d) \pm   D_{\lambda_d} (-z_d) \right]&
\textrm{for~} \vert z \vert < (\frac{m \omega }{\hbar})^{1/2} a
\\
\pm c_1 D_{\lambda} (z) & \textrm{for~} z > (\frac{m \omega
}{\hbar})^{1/2} a
\end{cases}
\label{totalWF}
\ee
where $+$ and $-$ signs stand for the even and odd eigenfunctions in
the second and third regions of the wave functions in Eq.
\eqref{totalWF}. Applying the continuity conditions on the wave
functions and their derivatives, at $\vert x \vert = a$, one can
express $c_2$ in terms of $c_1$ and get equations which determine
the eigenvalues of the Hamiltonian for the even and odd
eigenfunctions, respectively:
\bea
\sqrt{ \frac{\omega}{\omega_d}} \left[
D_{\lambda_d}(B)+D_{\lambda_d}(-B) \right] G_{\lambda}(A) &=& \left[
G_{\lambda_d}(B)- G_{\lambda_d}(-B) \right] D_{\lambda}(A)
\label{eigroote}
\\
\sqrt{ \frac{\omega}{\omega_d}} \left[
D_{\lambda_d}(B)-D_{\lambda_d}(-B) \right] G_{\lambda}(A) &=& \left[
G_{\lambda_d}(B)+G_{\lambda_d}(-B) \right] D_{\lambda}(A)
\label{eigrooto}
\\
\textrm{for ~} D_{\lambda}(A) \neq 0 & \textrm{and}&  G_{\lambda}(A)
\neq 0 , \nonumber
\eea
where
\be
G_{\lambda}(z)=\frac{d}{dz} D_{\lambda} (z),
\label{derivativeWF}
\ee
and $A=\sqrt{\frac{m \omega}{\hbar}} \; a$, $B=\sqrt{ \frac{m
\omega_d}{\hbar}} \; a$.
The Eqs. \eqref{eigroote} and \eqref{eigrooto}, corresponding to the
even and odd eigenfunctions respectively, have infinite number of
roots and can be solved numerically to determine $\lambda$ and
therefore the eigenvalues $E_{\lambda}=(\lambda+1/2) \hbar \omega$.


\section{JWKB approximation \label{JWKB} }

In the previous section we obtained analytically the eigenvalue Eqs.
\eqref{eigroote}, \eqref{eigrooto} and the eigenfunctions of the
Hamiltonian in Eq. \eqref{hamiltonian} whose potential term is given
in Eq. \eqref{potential}. However, to determine the eigenvalues one
needs to solve the Eqs. \eqref{eigroote} and \eqref{eigrooto}
numerically \footnote{We use Mathematica for the numerical solutions
of the Eq. \eqref{eigroote} and \eqref{eigrooto}.}. As $U_0$  i.e.
the depth of the potential increases, the eigenvalues of the low
lying states decrease. As the eigenvalues decrease one needs to
check the accuracy of the values obtained by numerical methods. For
this aim we apply JWKB \footnote{The JWKB approximation is called as
WKB approximation in most physics books. However the mathematical
formalism is introduced by H. Jeffreys \cite{jeffreys} so we call it
JWKB approximation.} approximation because it gives the exact
eigenvalues \cite{ghatak,huruska,landau,schiff} for harmonic
oscillator Hamiltonian whose potential term is one of the shape
invariant potentials \cite{huruska}.

The JWKB approximation method gives rise to the well known JWKB
quantization formula to find the eigenenergies  (e.g. see
\cite{schiff}),
\be
\int_{x_1}^{x_2} \left[ 2m \left(E_n-V(x) \right) \right]^{1/2}
dx=(n+\frac{1}{2}) \hbar
\label{JWKBquantization}
\ee
where $x_1$ and $x_2$ are classical turning points found by the
equality $E_n=V(x_{1,2})$. Since the potential function is different
for $\vert x \vert \leq a$ and $\vert x \vert > a$ the quantization
formula differs for $E_n < (1/2) m \omega^2 a^2$ and for $E_n> (1/2)
m \omega^2 a^2$. We will denote the eigenenergies whose values are
less than $ V(a)=(1/2) m \omega^2 a^2$ by $E_n^{(1)}$. For these
eigenenergies, Eq. \eqref{JWKBquantization} becomes
\bea
& & \int_{x_1}^{x_2} \left[ 2m \left(E_n^{(1)}-V(x) \right)
\right]^{1/2} dx  = \int_{x_1}^{x_2} \sqrt{ 2m \left[E_n^{(1)}-
\left(\frac{1}{2} m \omega^2 x^2-U_0
(1-\frac{x^2}{a^2}) \right) \right] }  dx =\nonumber \\
& &\int_{x_1}^{x_2} \sqrt{ 2m \left[E_n^{(1)}+U_0- \left(\frac{1}{2}
m \omega_d^2 x^2  \right) \right] }\; dx =(n+\frac{1}{2} )\hbar \pi
\label{Indimple}
\eea
where $\omega_d$ is defined in Eq. \eqref{severaldefs} and $x_{1,2}=
\pm \sqrt{(2 E_n^{(1)}+U_0)/(m \omega_d^2) }$. It is easy to solve
Eq. \eqref{Indimple} for $ E_n^{(1)}$:
\be
E_n^{(1)}=-U_0+(n+\frac{1}{2}) \hbar \omega_d \quad \textrm{for}
\quad n=0,1,\ldots,n'-1   \quad .
\label{eigeenindimple}
\ee
Here, ($n'$) is the number of the eigenstates  whose eigenvalues are
less than $V(a)$. It is necessary to determine the lowest level
index for the eigenstates in the upperlying region whose eigenvalues
are larger than $V(a)$ (See Eq. \eqref{Outdimple}). $n'$ is
determined by the formula
\be
n'=\lfloor \frac{\frac{1}{2} m \omega^2 a^2-E_0^{(1)}}{\hbar
\omega_d} \rfloor +1 ,
\label{eigeenindimpleno}
\ee
where the brackets denote the floor function.
\begin{table}
\caption{The eigenenergies of the first few eigenstates for
$\hbar=1, \; m=1/2, \; \omega=1, \; a=3, \; U_0=10$. For these
values $n'=5$ and $V(a)=2.25$. All the energies are in units of
$\hbar \omega$. } \label{om1w3a1U010}
\begin{tabular}[b]{|c|c|c|c|c|} \hline \multicolumn{5}{|c|} {The Eigenenergies} \\
\hline & The Eigenstate No  &  Analytic & JWKB & Difference
$\vert$Analytic-JWKB $\vert$\\ \hline &   0 & -8.8333& -8.8333& 0.0000 \\
\cline{2-5} & 1 & -6.5001 & -6.5000& 0.0001
\\ \cline{2-5} $ E_n^{(1)} $& 2& -4.1681 &-4.1667 & 0.0014 \\ \cline{2-5} & 3 & -1.8447 &-1.8333 &0.114 \\ \cline{2-5} & 4
&0.4355 &0.5000 &0.0645 \\  \hline & 5 &2.5432 & 2.6852& 0.1420\\ \cline{2-5} & 6 & 4.1711 &4.0504 &0.1207 \\
\cline{2-5} & 7 & 5.2756 &5.2638 & 0.0118\\ \cline{2-5} $ E_n^{(2)} $&8 & 6.3845&6.4181 & 0.0336\\\cline{2-5} & 9 &7.5823 & 7.5390& 0.0433\\
\cline{2-5}& 10 &8.6277 &8.6381 & 0.1040\\ \cline{2-5}& 11 &9.7171 &9.7217 & 0.0046\\
\hline
\end{tabular}
\end{table}
The JWKB quantization equation \eqref{JWKBquantization} for the
energy eigenvalues of the eigenstates in the upperlying region
becomes
\bea
& & 2 \bigg\{ \int_0^{a} \sqrt{ 2m \left[E_n^{(2)}+U_0-
\left(\frac{1}{2} m \omega_d^2 x^2  \right) \right] }\;  dx  +
\int_{a}^{x_2} \sqrt {2m
\left(E_n^{(2)}- \frac{1}{2} m \omega^2 x^2 \right)} \; dx \bigg\} \nonumber \\
&=&(n+\frac{1}{2} )\hbar \pi   \quad \textrm{for} \quad n=n',n'+1,
\ldots , \label{Outdimple}
\eea
where $x_2=\sqrt{(2 E)/(m \omega^2)}$ and $n'$ is given by Eq.
\eqref{eigeenindimpleno}. We take the integrals in Eq.
\eqref{Outdimple} only for positive values of $x$ utilizing the fact
that the potential function is an even function. After taking the
integrals the JWKB quantization equation for $E_n^{(2)}>V(a)$ gives
\bea
& & \frac{2 (E_n^{(2)}+U_0 )}{\omega_d} \arcsin\left[
\frac{a}{\sqrt{\frac{2 (E_n^{(2)}+U_0 )}{m \omega_d^2}}} \right] + m
a \omega_d \sqrt{\frac{2 (E_n^{(2)}+U_0 )}{m \omega_d^2}-a^2 }
\nonumber \\ &=& \frac{ E_n^{(2)}}{\omega} \bigg\{ \pi-  2\arcsin
\left[ \frac{a}{\sqrt{\frac{ 2 E_n^{(2)}}{m \omega^2}}} \right]
\bigg\} - m a \omega \sqrt{\frac{2 E_n^{(2)}}{m \omega^2}-a^2 } =
(n+\frac{1}{2}) \hbar  \pi .
\label{eigenoutdimple}
\eea
This equation also can be solved only numerically like Eq.
\eqref{eigroote} and \eqref{eigrooto}. However its solution is much
easier than the Eq. \eqref{eigroote} and \eqref{eigrooto} in which
one has to find the roots of infinite series.

We compare the values of the eigenenergies obtained by the JWKB
approximation and those found by the numerical solutions of the Eqs.
\eqref{eigroote} and \eqref{eigrooto} for different $U_0$ and $a $
values in the Tables \ref{om1w3a1U010} and \ref{realistic}. In the
Table \ref{om1w3a1U010}, we take $\hbar=1, \; 2m=1$. In the Table
\ref{realistic} (at the end of the manuscript), we take the true
value for $\hbar$ and more realistic values for the other parameters
\cite{kurn}. In these tables the first $n'$ eigenenergies, denoted
by $E_n^{(1)}$, are less than $V(a)$, and their JWKB values are
calculated by the Eq. \eqref{eigeenindimple} and $E_n^{(2)}$s are
larger than $V(a)$ and their JWKB values are calculated by Eq.
\eqref{eigenoutdimple}.
\begin{table}[ht!]
\caption{The eigenenergies of low lying eigenstates for $ \omega=2
\pi 20$Hz, $m=23$ amu, $a=11 \mu$m, \; $U_0=1.0\, 10^{-30}$J and
$\hbar$ is in SI units. For these values $n'=15$ and $V(a)=2.75 \,
\hbar \omega$. All the energies are in units of $\hbar \omega$.}
\label{realistic}
\begin{tabular}[b]{|c|c|c|c|c|}\hline \multicolumn{5}{|c|}{The Eigenenergies}
\\ \hline & The Eigenstate No  &  Analytic & JWKB & Difference
$\vert$(Analytic-JWKB)$\vert$  \\ \hline & 0 & -72.7948  & -72.7948 & 0.0000 \\
\cline{2-5} & 1 & -67.4650 & -67.4650&  0.0000
\\ \cline{2-5} &2 & -62.1353 & -62.1353 & 0.0000 \\ \cline{2-5}& 3 & -56.8055 & -56.8055 & 0.0000\\ \cline{2-5} &4
& -51.4758 & -51.4758 & 0.0000\\ \cline{2-5}& 5 & -46.1461 &
-46.1461 &
0.0000\\ \cline{2-5} &6 & -40.8163 & -40.8163 & 0.0000 \\
\cline{2-5} $E_n^{(1)} $&7 &-35.4866 & -35.4866 & 0.0000 \\
\cline{2-5}& 8 & -30.1570 & -30.1569 & 0.0001
\\ \cline{2-5}& 9 & -24.8278 & -24.8271 & 0.0007\\ \cline{2-5} &10 & -19.5005  & -19.4974 &
0.0031
\\ \cline{2-5} &11 & -14.1807 & -14.1676 & 0.0131 \\
\cline{2-5} &12& -8.8874 & -8.8379 & 0.0495
\\ \cline{2-5}& 13& -3.6816 & -3.5082 & 0.1734  \\ \cline{2-5} & 14& 1.2167 &1.8216 &
0.6049 \\ \hline &15 & 4.8125 & 4.4790 & 0.3335 \\ \cline{2-5} &16 & 6.0940  & 6.0210 & 0.0730\\
\cline{2-5} &17 &7.3188 & 7.4137  & 0.0949\\ \cline{2-5}& 18  & 8.8970  & 8.7300& 0.1670 \\ \cline{2-5}& 19 & 9.8991 &9.9980 & 0.0989\\
\cline{2-5} $E_n^{(2)} $ & 20 & 11.3352 & 11.2316 & 0.1036  \\
\cline{2-5} &21 & 12.4303 &
12.4396 & 0.0093 \\ \cline{2-5}& 22 & 13.6232 & 13.6273 & 0.0041 \\
\cline{2-5}  & \vdots & \vdots & \vdots & \vdots \\ \cline{2-5} &499
& 497.1836 &
497.1835 & 0.0001 \\ \cline{2-5} &500 & 498.1856 & 498.1857 & 0.0001 \\
\hline
\end{tabular}
\end{table}

The potential energy function given in Eq. \eqref{potential} changes
its form at $ \vert x \vert = a $ where its value is $V(a)$. Unless
the eigenenergies are close to this value, the JWKB approximation
gives accurate results. However, as the values of the eigenenergies
get close to the $V(a)$, the accuracy of the JWKB approximation
decreases. It is still useful to apply JWKB approximation in this
interval because in numerical calculations one may skip some of the
eigenstates since the difference between the successive eigenvalues
are not uniform in this interval. The reason that the JWKB
approximation gives accurate results for the ground and first few
excited eigenstates but not for the eigenstates whose eigenenergy
values are around $V(a)$ is the following: The eigenfunctions of the
ground and first few excited states go to zero very rapidly as
$\vert x \vert$ increases and they are almost zero at $\vert x \vert
=a$. So these eigenfunctions behave like they are the eigenfunctions
of the shifted harmonic oscillator potential $V(x)=-U_0 +(1/2)\,
m\omega_d^2 x^2$ for which the JWKB approximation gives exact
results. In other words they don't ``feel" the change of the form of
the potential. For the excited states whose eigenvalues are much
larger than $V(a)$, the JWKB approximation becomes again accurate.
We can explain this also using the eigenfunctions of these states.
These eigenfunctions are nonzero up to very large values compared to
$\vert x \vert =a$. So the region where the dimple potential exists
$\vert x \vert  \leq a$ is very narrow compared to the width of
these eigenfunctions. Therefore, they are not much effected by the
dimple potential.


\section{Transition Amplitudes \label{TA} }

In a recent paper, Garrett et. al. report the formation of a
Bose-Einstein condensate without cooling in an anisotropic harmonic
trap  using dimple potentials \cite{garrett}. In this study, they
investigate the behavior of a Bose gas in a harmonic trap with
``narrow" and ``wide" dimples both for adiabatic and sudden
processes. For the sudden turn of the dimple, the sudden
perturbation theory can be used to calculate the condensate
fraction. By describing the harmonic trap with a dimple potential by
Eq. \eqref{potential}, it is possible to apply sudden perturbation
theory since one can calculate the eigenfunctions for both case
(only harmonic trap $V(x)=(1/2) \, m \omega^2 x^2$ for $-\infty < x
< \infty$ and harmonic trap with a dimple potential).
\begin{figure}
\includegraphics{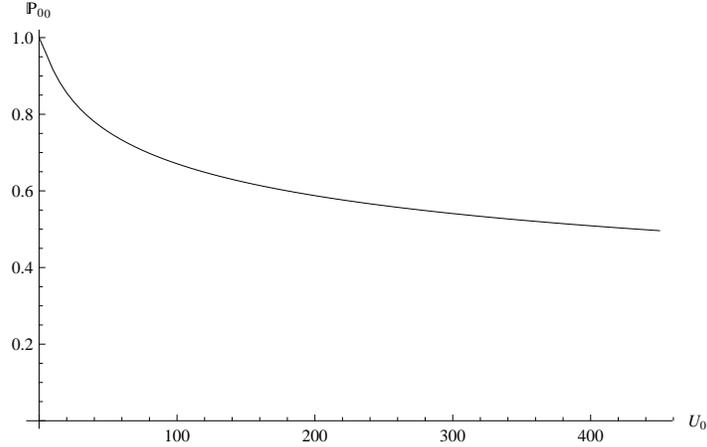}
\caption{The change of the transition probability from the ground
state of the harmonic oscillator Hamiltonian to the ground state of
the harmonic oscillator with a dimple potential $\mathbb{P}_{00}$
with respect to the depth of the dimple $U_0$.} \label{tafig1}
\end{figure}

In this section, we will present the sudden perturbation
calculations for one dimensional case. Before the turn on of the
dimple, the potential is only the harmonic potential and the
eigenfunctions for this case are \cite{landau,schiff}:
\be
\psi_n{(z)}=(\frac{ (m \omega)^{1/2} }{(\pi \hbar)^{1/2}
{2^{n}}{n!}})^{1/2} \exp({-z^2/2})\; H_n(z) \, ,
\label{eigenfunctionh}
\ee
where $z $ is given by Eq. \eqref{dimenionless}.
The eigenfunctions of the potential given by Eq. \eqref{potential}
are found in Section \ref{HTWPP} and given in Eq. \eqref{totalWF}.
Using the sudden perturbation theory \cite{landau}, we calculate the
transition amplitudes as
\be
t_{n \lambda}(z)= \sqrt{\frac{\hbar}{m \omega} }
\int_{-\infty}^{\infty} \psi_n (z)\, \Psi_{\lambda} (z) \, dz.
\label{transamp}
\ee
In this equation one has to choose $c_1$ in Eq. \eqref{totalWF} such
that $ \Psi_{\lambda} (z)$ is normalized to unity to find the
transition probabilities as:
\be
P_{n \lambda}=\vert t_{n \lambda} \vert^2.
\label{transprob}
\ee
\begin{figure}[b!]
\includegraphics{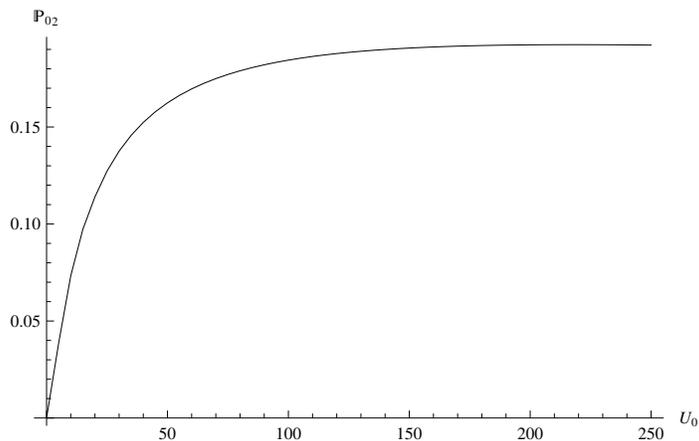}
\caption{The change of the transition probability from the second
excited state of the harmonic oscillator Hamiltonian to the ground
state of the harmonic oscillator with a dimple potential
$\mathbb{P}_{02}$ with respect to the depth of the dimple $U_0$.}
\label{tafig2}
\end{figure}

We present the change of the transition probabilities with respect
to $U_0$ from the ground state and from the second excited state of
the harmonic oscillator to the ground state of the harmonic
oscillator with a dimple for the values $\hbar=1, \; m=1/2, \;
\omega=1, \; a=3$ in the Figures \ref{tafig1} and \ref{tafig2},
respectively. The transitions from the even eigenstates to odd
eigenstates are zero since the wave functions of these states are
even and odd functions.

In following studies we aim to use the truncated parabolic potential
 to model dimple potentials used in
Bose Einstein condensation experiments. Similar transition
probability calculations from higher levels to the new ground state
can be used for the calculation of the condensate fraction of a BEC
in a harmonic trap with a dimple when the dimple is turned on
suddenly \cite{garrett}.
\section{Dirac $\delta$ potential as a limiting case of the Parabolic Potential \label{DDPaLC}}

Recent papers show that modeling dimple potentials by Dirac $\delta$
functions reveal basic properties of a BEC in a harmonic trap with a
dimple \cite{utem,busch}. In this section we will present that the
$\delta$ potential is included in the truncated parabolic potential
as a special case.

We will first show that the function
\be
f_{\delta }(x)= \begin{cases} \frac{3}{4 a} \left( 1-\frac{x^2}{a^2}
\right) & \textrm{for~} \vert x \vert \leq a \\ 0 &  \textrm{for~}
\vert x \vert
> a
\end{cases}
\label{deltadist}
\ee
represents the Dirac $\delta$ function. There is an advantage of
this representation compared to other representations in the
literature (see e.g.  \cite{cohen} Appendix II). When the function
$f_{\delta}(x)$ in Eq. \eqref{deltadist} is used as a potential for
a Hamiltonian like in Eq. \eqref{potential}, the resulting
time-independent \Sch is analytically solvable.

It is clear that the representation given in Eq. \eqref{deltadist}
satisfies the condition $ \int_{-\infty}^{\infty} \delta(x) dx=1$
used in defining  Dirac $\delta$ function. We will show that this
definition satisfies also the sampling property of $\delta$ function
\cite{hoskins}: $\int_{-\infty}^{\infty} \delta(x)  h(x) dx=h(0) $
for $a\rightarrow 0 $ where $h(x)$ denotes an analytical function.
That is
\be
\lim_{a \to 0} \int_{-\infty}^{\infty} f_{\delta }(x) h(x) dx
=\lim_{a \to 0} \int_{-a}^{a} f_{\delta }(x) h(x) dx= h(0) .
\label{deltacondition}
\ee
For this  aim we define the functions $H_1(x)$, $H_2(x)$, $H_3(x)$
as \footnote{The fact that $H_1(x)$, $H_2(x)$ and $H_3(x)$ are not
unique do not affect the result.}
\be
\frac{d H_1(x)}{dx}=h(x), \qquad  \frac{d^2 H_2(x)}{dx^2}=h(x)
\qquad \frac{d^3 H_3(x)}{dx^3}=h(x).  \label{quadrature}
\ee
We apply three times integration by parts to the expression $
\lim_{a \rightarrow 0}\int_{-a}^{a} f_{\delta }(x) h(x) dx $ and get
\be
\lim_{a \rightarrow 0}\int_{-a}^{a} f_{\delta }(x) h(x) dx=\lim_{a
\rightarrow 0} \left\{ \frac{3}{2 a^2} \left[ H_2(a)+H_2(-a) \right]
- \left[ H_3(a) -H_3(-a) \right] \right\}.
\label{Deltaparts}
\ee
Using the Maclaurin series expansion of $H_2(a)$,  $H_2(-a)$ and
$H_3(a)$,  $H_3(-a)$ up to $a^3$ one can see that the right hand
side of the Eq. \eqref{Deltaparts} is equal to $h(0)$ in the limit
$a \rightarrow 0$.

It is well known that there is only one bound state of the Dirac
delta potential (see e.g. \cite{demiralp})
\be
 V_{\delta}=- \frac{\hbar^2}{2m} \, \sigma  \, \delta(x) .
\label{deltapot}
\ee
The solution of the eigenvalue equation for this potential is
$\kappa=\sigma/2$ where $\kappa=\sqrt{-(2 m E)}/\hbar \,$. Therefore
one finds \cite{demiralp}
\be
E=- \frac{\hbar^2 \sigma^2}{8m}.
\label{eigvalofdelta}
\ee
Here the coupling coefficient of the $\delta$ function is taken as
$-\hbar^2 \sigma /(2m)  $ for calculational convenience. We will now
calculate the bound state eigenvalues for the potential
\be
V(x)= \begin{cases} -U_0 \left( 1-\frac{x^2}{a^2} \right) &
\textrm{for~} \vert x \vert \leq a \\ 0 & \textrm{for~} \vert x
\vert >a
\end{cases}
\label{deltapotmod}
\ee
and present that for small $a$ the eigenvalue of the ground state
energy of this potential approximates to the ground state energy of
Dirac $\delta$ potential given in Eq. \eqref{deltapot}. The even and
odd eigenfunctions of the bound states for the potential  given in
Eq. \eqref{deltapotmod} is:
\be
\psi_e(x)= \begin{cases} A_e e^{\kappa x} & \textrm{for~}  x < -a
\\
B_e \left[D_{\gamma_d} (\sqrt{\frac{ m \nu }{\hbar} } x) +
D_{\gamma_d} (-\sqrt{\frac{ m \nu }{\hbar}} x) \right]
 & \textrm{for~} \vert x \vert \leq a
\\ A_e e^{-\kappa x} & \textrm{for~}  x >a
\label{evendeltaWF}
\end{cases}
\ee
and
\be
\psi_o(x)= \begin{cases} A_o e^{\kappa x} & \textrm{for~}  x < -a
\\
B_o \left[D_{\gamma_d} (\sqrt{\frac{ m \nu }{\hbar}} x) -
D_{\gamma_d} (-\sqrt{\frac{ m \nu }{\hbar}} x)\right]
 & \textrm{for~} \vert x \vert \leq a
\\ -A_o e^{-\kappa x} & \textrm{for~}  x > a
\end{cases}
\label{oddeltaWF}
\ee
where $\nu=\sqrt{2U_0/(m a^2)}$, $\gamma_d=(E+U_0)/(\hbar \nu)-1/2$
and the coefficients $A_e$, $A_o$, $B_e$ and $B_o$ are to be
determined from normalization and continuity conditions of the wave
function and its derivative. Applying these continuity conditions
one gets the eigenvalue equation for the potential  given in Eq.
\eqref{deltapotmod} of the even and odd eigenfunctions, respectively
as:
\be
- \kappa \sqrt{\frac{\hbar}{m \nu}}=-\sqrt{- (2 \gamma+1)}=
\frac{G_{\gamma_d}(\sqrt{\frac{m \nu}{\hbar} } a )
-G_{\gamma_d}(-\sqrt{\frac{m \nu}{\hbar} }a )
}{D_{\gamma_d}(\sqrt{\frac{m \nu}{\hbar} }a ) +
D_{\gamma_d}(-\sqrt{\frac{m \nu}{\hbar} }a ) }
\label{eigenvalueODimplee}
\ee
\be
- \kappa \sqrt{\frac{\hbar}{m \nu}}=-\sqrt{- (2 \gamma+1)}=
\frac{G_{\gamma_d}(\sqrt{\frac{m \nu}{\hbar} } a )
+G_{\gamma_d}(-\sqrt{\frac{m \nu}{\hbar} }a )
}{D_{\gamma_d}(\sqrt{\frac{m \nu}{\hbar} }a ) -
D_{\gamma_d}(-\sqrt{\frac{m \nu}{\hbar} }a ) }
\label{eigenvalueODimpleo}
\ee
where $\gamma=E/(\hbar \nu)-1/2$; $D_{\lambda}(z)$ and
$G_{\lambda}(z)$ are defined in Eqs. \eqref{hypergeomsol} and
\eqref{derivativeWF}, respectively. By taking into account the
definition of the representation for $\delta$ in Eq.
\eqref{deltadist}, the comparison of Eqs. \eqref{deltapot} and
\eqref{deltapotmod} reveals that $\sigma=8 m a U_0/(3 \hbar^2)$. As
$a \rightarrow 0$, for constant $U_0 a $, the eigenvalue Eq.
\eqref{eigenvalueODimplee} reduces to the Eq. \eqref{eigvalofdelta}
which gives the bound state eigenvalue of the $\delta$ potential.
One can show this by expanding the right hand side of the Eq.
\eqref{eigenvalueODimplee} into a Taylor series. For constant $U_0
a$, the left hand side of the equality in Eq.
\eqref{eigenvalueODimplee} is order of $a^{3/4}$ \footnote{Since we
take $U_0 a$ is equal to a constant c, $\nu=\sqrt{2U_0/(m a^2)}$
becomes  $\nu=\sqrt{2c/(m a^3)}$. Hence $\nu$ is of order $a^{-3/2}
$. }. Therefore, we expand the right hand side in terms of $\sqrt{m
\nu/\hbar} \; a $ so that the expansion terms in the right hand side
include terms up to $a^{3/4}$. Doing this, we get the equality
\be
 -\sqrt{-(2 \gamma+1)}= -(2 \gamma_d +1) \sqrt{\frac{m
\nu}{\hbar} } a+\frac{7}{6} \left( \sqrt{\frac{m \nu}{\hbar} } a
\right)^3 .
\label{reducedeig}
\ee
Substituting $\nu=\sqrt{2U_0/(m a^2)}$, $\gamma_d=(E+U_0)/(\hbar
\nu)-1/2$ and $\gamma=E/(\hbar \nu)-1/2$ into this equation we find
\be
E=-\frac{8}{9} \frac{m U_0^2 a^2 }{\hbar^2}.
\label{energyequality}
\ee
Since $\sigma=8 m a U_0/(3 \hbar^2)$, Eq. \eqref{energyequality} is
identical to the Eq. \eqref{eigvalofdelta}, which reveals that as $a
\rightarrow 0$ the representation given in Eq. \eqref{deltadist}
gives the same ground state eigenvalue with $\delta$ potential for
the solution of the Schr\"{o}dinger equation. Doing a similar
expansion to the Eq. \eqref{eigenvalueODimpleo}, we see that this
equation does not have a solution in the limit $a\rightarrow 0 $.
That means there is no odd bound state which proves there is only
one bound state for the potential given in Eq. \eqref{deltapotmod}
in the limit $a\rightarrow 0 $.

We also calculate the reflection and tunneling probabilities for the
truncated parabolic potential given in Eq. \eqref{deltapotmod} for
the scattering states. Then, we show that as $a \rightarrow 0$,  the
tunneling and reflection amplitudes for this potential reduce to the
transition and reflection amplitudes for the Dirac $\delta$
potential given in Eq. \eqref{deltapot}. The wave function for a
scattering state of a $\delta$ potential incoming from $-\infty$
with an energy $E>0$ can be written as:
\be
\psi_{\delta s}(x)=\begin{cases} e^{ i k a x}+ R_{\delta} e^{- i k a
x} & \textrm{for~} x < -a
\\
T_{\delta} e^{i k a x} & \textrm{for~}  x > a
\end{cases}
\label{scattwavedelta}
\ee
where $k=\sqrt{2mE/\hbar^2}$ and we take the coefficient of the wave
incoming from $-\infty$ equal to unity for calculational
convenience. The transition and reflection amplitudes for this
potential is easily found to be
\bea
T_{\delta}&=&\frac{2 i k}{2 i k+\sigma} \label{tunneldelta}
\\
R_{\delta} &=& - \frac{\sigma}{2 i k + \sigma} \; .
\label{reflectdelta}
\eea
The wave function for a scattering state incoming from $-\infty$
with an energy $E>0$ for the potential in Eq. \eqref{deltapotmod}
can be written as:
\begin{equation}
\psi_s(x)=\begin{cases} e^{ i k a x}+ R e^{- i k a x} &
\textrm{for~} x < -a
\\
c_1 D_{\gamma_d} (\sqrt{\frac{ m \nu }{\hbar}} x) + c_2 D_{\gamma_d}
(-\sqrt{\frac{ m \nu }{\hbar}} x) & \textrm{for~} \vert x \vert \leq
a
\\T e^{i k a x} & \textrm{for~}  x > a \, .
\end{cases}
\label{scattwave}
\end{equation}
The norm squares of $R $ and $T $ given in Eq. \eqref{scattwave}
give the probability of reflection and tunneling, respectively.
These coefficients are calculated using the continuity of the wave
function and its derivative at points $x=\vert a \vert$. In order to
present the result for $R$, we define
\be
F_1(\lambda,a)={-\Phi}\left(-\frac{\lambda }{2},\frac{1}{2},a^2
s^2\right)+2 (1+\lambda ) \Phi\left(-\frac{\lambda
}{2},\frac{3}{2},a^2 s^2\right)
\label{function1}
\ee
\be
F_2(\lambda,a)=\left(k^2+s^2+a^2 s^4\right) \Phi\left(-\frac{\lambda
}{2},\frac{1}{2},a^2 s^2\right)-2 s^2 \left(1+a^2 s^2\right)
(1+\lambda ) \Phi\left(-\frac{\lambda }{2},\frac{3}{2},a^2
s^2\right)
\label{function2}
\ee
\be
F_3(\lambda,a)= 3 \left(-1+i a k+a^2 s^2\right)
\Phi\left(\frac{1-\lambda }{2},\frac{3}{2},a^2 s^2\right)+2 a^2 s^2
(-1+\lambda ) \Phi\left(\frac{3-\lambda }{2},\frac{5}{2},a^2
s^2\right)
\label{function3}
\ee
\be
F_4(\lambda,a)= 2 a s^2 \lambda \Phi\left(1-\frac{\lambda
}{2},\frac{3}{2},a^2 s^2\right)+\left(i k+a s^2\right)
\Phi\left(-\frac{\lambda }{2},\frac{1}{2},a^2 s^2\right) ,
\label{function4}
\ee
where $\Phi(\alpha,\gamma;y)$ is the confluent hypergeometric
function \cite{lebedev}. In terms of the functions $F_1(\lambda,a)$
to $F_4(\lambda,a)$ the reflection amplitude is
\be
R=-\frac{a  e^{-2 i a k} \left[2 a^2 s^4 (2+\lambda )
\Phi\left(\frac{1-\lambda }{2},\frac{5}{2},a^2 s^2\right)
F_1(\lambda,a)+  3 \Phi\left(\frac{1-\lambda} {2},\frac{3}{2},a^2
s^2\right) F_2(\lambda,a)\right] }{F_3(\lambda,a) \; F_4(\lambda,a)
} \, .
\label{reflection}
\ee
In order to demonstrate $T$, we additionally define
\be
F_5(\lambda,a)= -2 a^2 s^2 (2+\lambda)
\Phi\left(\frac{1}{2}-\frac{\lambda }{2},\frac{5}{2},a^2 s^2\right)
\Phi\left(-\frac{\lambda }{2},\frac{1}{2},a^2 s^2\right)
\label{function5}
\ee
\be
F_6(\lambda,a)=\Phi\left(-\frac{\lambda }{2},\frac{1}{2},a^2
s^2\right)+2 a^2 s^2 (1+\lambda ) \Phi\left(-\frac{\lambda
}{2},\frac{3}{2},a^2 s^2\right).
\label{function6}
\ee
In terms of the functions $F_3(\lambda,a)$ to $F_6(\lambda,a)$
\be
T=-\frac{i e^{-2 i a k} k \left[F_5(\lambda,a)+ 3
\Phi\left(\frac{1-\lambda} {2},\frac{3}{2},a^2 s^2\right)
F_6(\lambda,a) \right] }{F_3(\lambda,a) \; F_4(\lambda,a) } \, .
\label{tunneling}
\ee
In the appendix, we show the change of the tunneling and reflection
probabilities with respect to energy $E$, width $a$ and depth $U_0$
of the truncated parabolic potential. Now we will show that the
tunneling amplitude $T$ for the potential given in Eq.
\eqref{deltapotmod} goes to $T_{\delta}$ in Eq. \eqref{tunneldelta}.
For this reason assuming $U_0 a$ is constant, we expand the
numerator and the denominator of the right hand side of the Eq.
\eqref{tunneling} with respect to $a$ and since we are interested in
the limit $a\rightarrow 0 $, we keep only the constant terms which
are independent of $a$. Doing this, we find
\be
\lim_{a \to 0} T =\frac{3 i k}{3ik+4mU_0 a/\hbar^2}.
\ee
Since $\sigma=8 m a U_0/(3 \hbar^2)$, we see that $T$ reduces to
$T_\delta$ given in Eq. \eqref{tunneldelta} for constant $U_0 a$.

Finally, we show that the eigenvalues of the Hamiltonian with the
potential given in Eq. \eqref{potential} approximates to the
eigenvalues of the Hamiltonian with the potential
\be
V(x)=\frac{1}{2} m \omega^2 x^2- \frac{\hbar^2}{2m} \sigma \delta(x)
\label{harmplusdelta}
\ee
for $U_0=3 \hbar^2 \sigma/(8 m a)$ as $a \rightarrow 0$. The
eigenvalues of even states of the Hamiltonian with the potential of
the Eq. \eqref{harmplusdelta} is equal to $E_{\lambda}=(\lambda+1/2)
\hbar \omega$ where $\lambda$ values are the roots of the equation
\cite{utem,avakian}
\be
\frac{\Gamma(\frac{1-\lambda}{2})}{\Gamma(-\frac{\lambda}{2})}=\frac{\Lambda}{4}.
\label{diraceigval}
\ee
Here $\Lambda=\sigma \sqrt{\hbar/(m \omega)}$. The odd eigenvalues
are equal to the odd eigenvalues of the harmonic oscillator
potential i.e. they are equal to $E_n=(n+1/2) \hbar \omega $ for
$n=1,3,5,\cdots $ \cite{utem}. As $a \rightarrow 0$, for constant
$U_0 a=c$, Eq. \eqref{eigroote} reduces to Eq. \eqref{diraceigval}
and the roots $\lambda$ of Eq. \eqref{eigrooto} go to
$\lambda_n=2n+1$ where $n=0,1,2,3, \cdots$. In order to show this,
we first expand $\sqrt{\omega/\omega_d}$ with respect to $a$ and
find
\be
\sqrt{\frac{\omega}{\omega_d} }=    \left( \frac{m a^3 \omega^2}{2c}
 \right)^{1/4} \left( 1+ \frac{m a^3 \omega^2}{2c}
 \right)^{-1/4}   \approx \;   \left( \frac{m a^3 \omega^2}{2c}
 \right)^{1/4} \left( 1-\frac{1}{4} \frac{m a^3 \omega^2}{2c}+ O(a^3) \right)
\label{omegafrac}
\ee
for $U_0 a=c$. Then we expand $D_{\lambda_d}(B)$,
$D_{\lambda_d}(-B)$  , $G_{\lambda_d}(B)$, $G_{\lambda_d}(-B)$ in
terms of $B=\sqrt{m \omega_d/ \hbar} \; a $; $G_{\lambda}(A)$ and
$D_{\lambda}(A)$ in terms of $A=\sqrt{m \omega/ \hbar}\; a $  in Eq.
\eqref{eigroote} and obtain for the eigenvalue equation of the even
eigenfunctions
\be
-\frac{8}{3} \; \frac{ \left( \Gamma(1/2)
\right)^2}{\Gamma(\frac{1-\lambda_d}{2})\Gamma(\frac{1-\lambda}{2})}\;
\frac{(m c)^{3/4}}{2^{1/4}\hbar^{3/2}} \; a^{3/4}=
2\left(\frac{m}{2c}\right)^{1/4}\omega^{1/2} \frac{
\Gamma(1/2)\Gamma(-1/2)}{ \Gamma(\frac{1-\lambda_d}{2})
\Gamma(-\frac{\lambda}{2})}\; a^{3/4}.
\ee
After rearranging the terms we get
\be
\frac{2}{3}\;\frac{U_0\, a\,
m^{1/2}}{\hbar^{3/2}\,\omega^{1/2}}=\frac{\Gamma(\frac{1-\lambda}{2})}{\Gamma(-\frac{\lambda}{2})}.
\label{paraboliceigene}
\ee
When we insert $\sigma=8 m a U_0/(3\hbar^{2})$ and $\Lambda=\sigma
\sqrt{\hbar/(m \omega)}$ into Eq. \eqref{paraboliceigene}, we see
that this equation reduces to the Eq. \eqref{diraceigval}.
Applying a similar procedure to the eigenvalue Eq. \eqref{eigrooto}
of the odd eigenfunctions, we see that it reduces to
\be
\frac{ \Gamma(-1/2)\Gamma(1/2)}{ \Gamma(\frac{1-\lambda}{2})
\Gamma(-\frac{\lambda_d}{2})}=0
\label{paraboliceigeno}
\ee
as $a \rightarrow 0$ for constant $U_0 a$. In this limit
$\lambda_d=-1/2$, therefore the right hand side of Eq.
\eqref{paraboliceigeno} can be equal to zero if and only if
$\Gamma(\frac{1-\lambda}{2})= \infty $. Since $\Gamma(-n)=\infty$
when $n=0,1,2,\cdots$, we get for $\lambda$
\be
\frac{1-\lambda}{2}=-n \quad  \Rightarrow \lambda=2n+1  \quad
\textrm{for~} n=0,1,2,\cdots \, .
\ee
Therefore, similar to the Dirac $\delta$ case, in the limit $a
\rightarrow 0$ the odd eigenvalues of the potential given in Eq.
\eqref{potential} reduces to the odd eigenvalues of the harmonic
oscillator potential with a Dirac $\delta$ potential at the origin.

The fact that the eigenvalues of the harmonic potential with a
symmetric truncated parabolic potential around the origin reduce to
the eigenvalues of the harmonic potential with a Dirac $\delta $ at
the origin, in the limit $a \rightarrow 0 $ for fixed $U_0 a$, shows
that the results obtained modeling the dimple potential with a Dirac
$\delta$  for a non-interacting BEC in a harmonic trap with a dimple
\cite{utem,busch} is included in the parabolic model of the dimple
as a special case.
\section{Conclusion \label{CONC}}

We propose, in this study, that the parabolic potential defined in
Eq. \eqref{potential} is more appropriate for modeling the dimple
potentials than the Dirac $\delta$ potential used in the literature.
Therefore, we showed that the parabolic potential includes the Dirac
$\delta$ potential as a special case and the wave functions,
eigenvalues, tunneling and reflection coefficients of the Dirac
$\delta$ potential can be obtained from those of the parabolic
potential as a limiting case.

In section \ref{HTWPP},  we have summarized the solution of the
Schr\"{o}dinger equation for a harmonic trap with a truncated
parabolic potential in one dimension and obtained eigenvalue
equations and eigenfunctions. Then, we presented the numerical
solutions of the eigenvalue equations. When the depth of the
parabolic potential increases the eigenenergies of the ground and
first few excited states decrease to more negative values as shown
in Table \ref{realistic}. By plotting the eigenfunctions of these
low lying states we have realized that as the eigenenergies decrease
the numerical solutions become instable. Therefore, in Section
\ref{JWKB}, we applied JWKB approximation to check the validity of
the numerically found eigenvalues. The results in Tables
\ref{om1w3a1U010}-\ref{realistic} show that the JWKB approximation
and the numerical solution for the eigenenergies agree very well
except in the transition region of the potential given in Eq.
\eqref{potential}. So, we conclude that one can use JWKB
approximation to find the eigenvalues for this potential when the
numerical solutions fail.

In section \ref{TA}, we obtained the formula for the transition
amplitudes and found the transition probabilities from the ground
and second excited eigenstates of the harmonic trap to the ground
state of the harmonic trap with a dimple described by the truncated
parabolic potential for different dimple depths, using the sudden
perturbation theory. In the following studies we aim to use these
and similar results in three dimensions for modeling the sudden turn
on of dimple potential experiments \cite{garrett}.

In section \ref{DDPaLC}, we have first shown that the truncated
parabolic function provides a representation of the Dirac $\delta$
function. Moreover, we have also demonstrated that the bound state
spectrums of the potentials given in Eqs. \eqref{deltapotmod} and
\eqref{potential} reduce to the spectrum of the Dirac $\delta$
potential and harmonic potential with a Dirac $\delta$  as $a
\rightarrow 0$, for fixed $U_0 a$. As one can see from the
Bose-Einstein distribution $\langle n_i \rangle =1/ (e^{\beta
(E_i-\mu)} -1)$  for a non-interacting Bose gas the effect of the
trapping potential to the thermodynamic properties of the gas comes
only through the eigenvalues of the potential. So, we come to the
following conclusion: The fact that the eigenvalues of the truncated
parabolic potential reduce to the eigenvalues of the Dirac $\delta $
potential in the limit $a\rightarrow 0$, for fixed $U_0 a$, shows
that the results obtained modeling the dimple potential with a Dirac
$\delta$ for a non-interacting Bose-Einstein condensate in a
harmonic trap with a dimple \cite{utem,busch} is included in the
parabolic model of the dimple as a special case.

\acknowledgments M\c{C}A and HU acknowledge the support by TUBITAK
(Project No:108T003). We also thank gratefully to Fatih Erman for
useful discussions.

\section{Appendix}
\begin{figure}[t!]
\includegraphics[height=8cm,width=15cm]{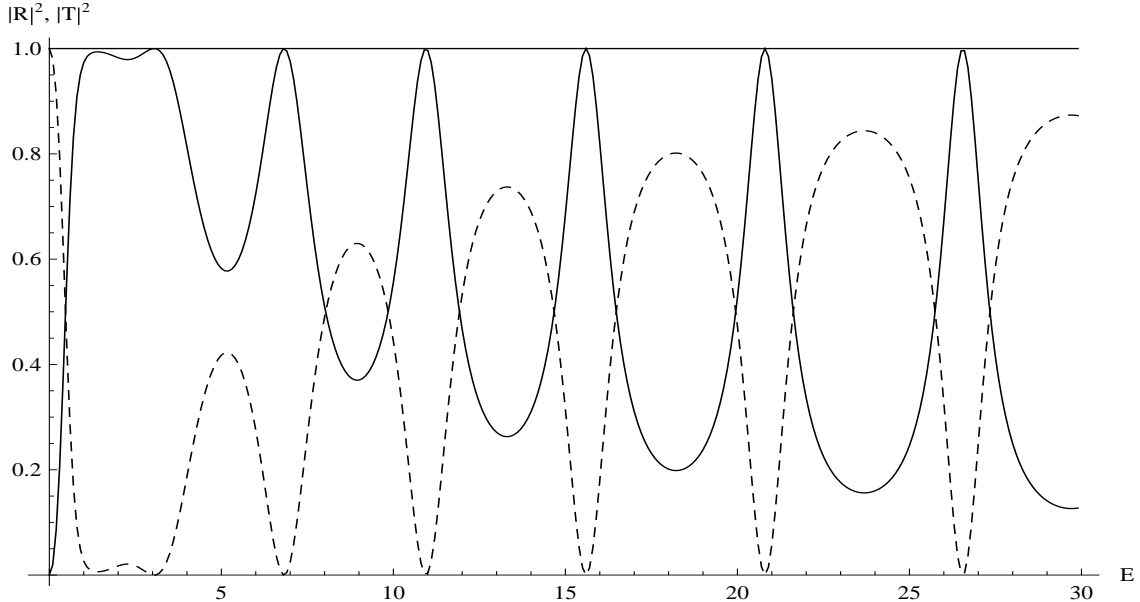}
\caption{The change of the tunneling and reflection probabilities
$\vert T \vert^2$ and $\vert R \vert^2$ for the truncated parabolic
potential,  Eq. \eqref{deltapotmod}, with increasing energy of the
incoming wave and for fixed $a=3$, $U_0=10$ in natural units. The
solid and dashed lines represent $\vert T \vert^2$ and $\vert R
\vert^2$, respectively.} \label{TRprobVE}
\end{figure}
We present in this appendix, the change of the tunneling and
reflection probabilities with respect to energy $E$ of the incoming
wave, width $a$ and depth $U_0$ of the potential given in Eq.
\eqref{deltapotmod}.  The scattering wave function, reflection ($R$)
and tunneling ($T$) amplitudes for this potential is given in Eqs.
\eqref{scattwave}, \eqref{reflection} and \eqref{tunneling},
respectively.
\begin{figure}[h!]
\includegraphics[height=8cm,width=15cm]{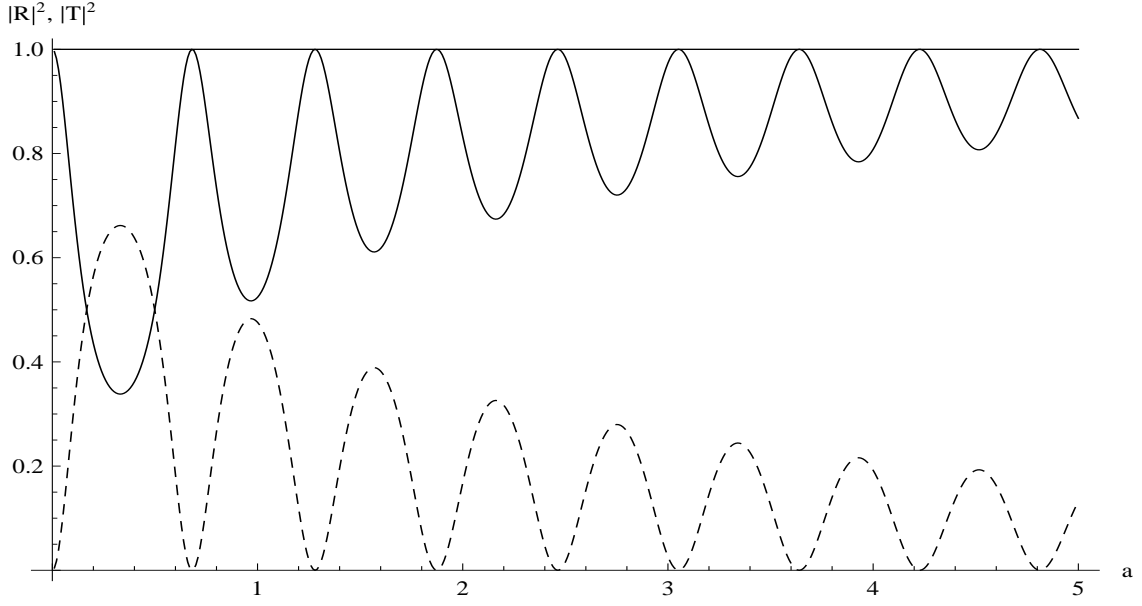}
\caption{The change of $\vert T \vert^2$ and $\vert R \vert^2$ for
the truncated parabolic potential,  Eq. \eqref{deltapotmod}, with
increasing potential width $a$ and for fixed $E=1$, $U_0=10$ in
natural units. The solid and dashed lines represent $\vert T
\vert^2$ and $\vert R \vert^2$, respectively.} \label{TRprobVa}
\end{figure}
\begin{figure}[t!]
\includegraphics[height=8cm,width=15cm]{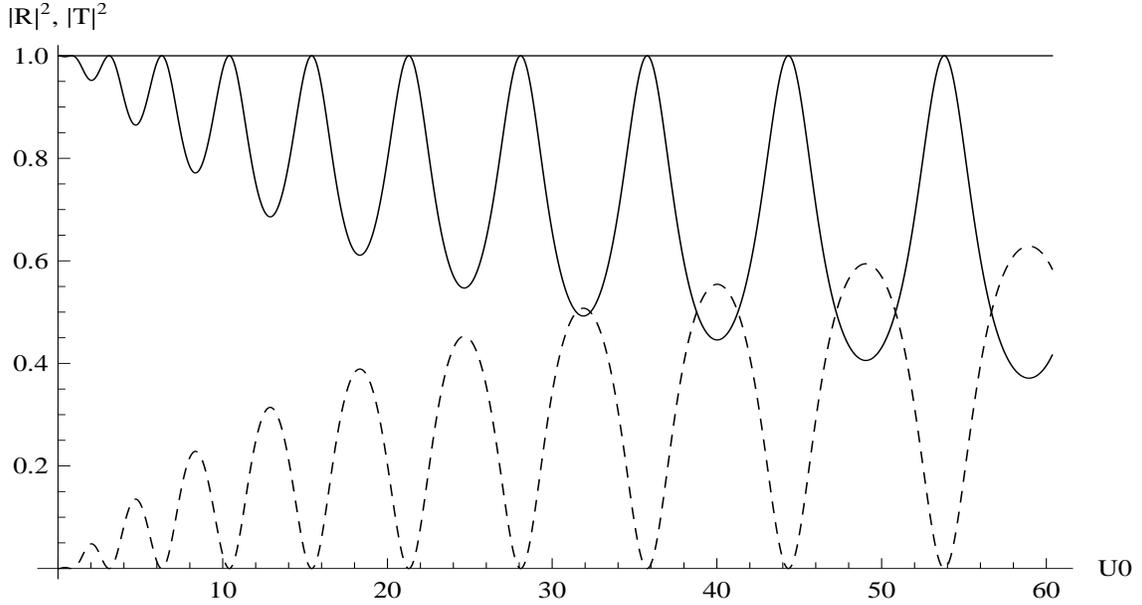}
\caption{The change of $\vert T \vert^2$ and $\vert R \vert^2$ for
the truncated parabolic potential,  Eq. \eqref{deltapotmod}, with
increasing potential depth $U_0$ and for fixed $E=1$, $a=3$ in
natural units. The solid and dashed lines represent $\vert T
\vert^2$ and $\vert R \vert^2$, respectively. } \label{TRprobVU0}
\end{figure}
We first present the change of the tunneling and reflection
probabilities with increasing energy $E$ of the incoming wave in
Fig. \ref{TRprobVE}, for fixed potential width $a=3$ and depth
$U_0=10$. Here we use natural units $\hbar=1$, $m=1/2$. One can see
from the Fig. \ref{TRprobVE}, as E increases from zero, the
tunneling probability first increases almost up to certainty.
However, then it begins to oscillate with increasing amplitudes and
widths such that local maximums are unity. Although the local
minimums of these oscillations decrease almost up to zero, they
never become equal to zero up to $E=1000$.

Then, in Fig. \ref{TRprobVa}, we show the variation of the tunneling
and reflection amplitudes for increasing $a$ and fixed $U_0=10$,
$E=1$. We see that the tunneling probability oscillates such that
local maximums touch one and as $a$ increases the amplitude of the
oscillations decrease. Finally, we demonstrate the change of the
$\vert T \vert^2$ and $\vert R \vert^2$ for increasing $U_0$ but for
fixed  $E=1$ and $a=3$ in Fig. \ref{TRprobVU0}. Similar to the case
of varying $E$, with increasing $U_0$, $\vert T \vert^2$ oscillates
with increasing amplitudes where local maximums are again one.


\end{document}